\documentclass[cits]{PoS}

\usepackage{textcomp}
\usepackage[latin1]{inputenc}
\usepackage{amsfonts}
\usepackage{amssymb}
\usepackage{amsbsy}
\usepackage{amsmath}
\usepackage{xspace}
\usepackage{graphicx}

\newcommand{\invfb}{\,{\rm fb}^{-1}}

\newcommand{\gev}{{\hbox{GeV}}}
\newcommand{\mmev}{{\hbox{MeV}/c^2}}

\newcommand{\mbc}{{M_{\textrm{bc}}}}
\newcommand{\deltae}{{\Delta E}}

\newcommand{\lint}{{L_{\textrm{int}}}}
\newcommand{\trho}{\theta_{\rho^+}}
\newcommand{\tdsst}{\theta_{\dsst}}
\newcommand{\crho}{\cos\trho}
\newcommand{\cdsst}{\cos\tdsst}
\newcommand{\ssqdsst}{\sin^2\tdsst}
\newcommand{\csqrho}{\cos^2\trho}
\newcommand{\csqdsst}{\cos^2\tdsst}
\newcommand{\ssqrho}{\sin^2\trho}
\newcommand{\dG}{\Delta\Gamma_s^{\rm CP}}
\newcommand{\G}{\Gamma_s}
\newcommand{\rfphi}{R_{f/\phi}}

\newcommand{\BR}{{\mathcal B}}

\newcommand{\FourS}{\Upsilon(4S)}
\newcommand{\FiveS}{\Upsilon(5S)}
\newcommand{\bs}{{B_s^0}}
\newcommand{\bsst}{{B_s^{\ast}}}
\newcommand{\barbsst}{{{\bar B_s}^{\ast}}}
\newcommand{\bsST}{{B_s^{(\ast)}}}
\newcommand{\ds}{{D_s^-}}
\newcommand{\dsst}{D_s^{\ast-}}

\newcommand{\KS}{{K_S^0}}

\newcommand{\jpsi}{{J\!\!/\!\!\psi}}
\newcommand{\fz}{f_0(980)}

\newcommand{\bsSTbsST}{{\bsST{\bar B_s}^{(\ast)}}}
\newcommand{\dsSTdsST}{D_s^{(\ast)+}D_s^{(\ast)-}}
\newcommand{\dsstdsst}{D_s^{\ast+}D_s^{\ast-}}
\newcommand{\dsstds}{D_s^{\ast\pm}D_s^{\mp}}
\newcommand{\dsds}{D_s^{+}D_s^{-}}

\newcommand{\bsdspi}{{\bs\to D_s^-\pi^+}}
\newcommand{\bsdsstpi}{{\bs\to D_s^{\ast-}\pi^+}}
\newcommand{\bsdsSTpi}{{\bs\to D_s^{(\ast)-}\pi^+}}

\newcommand{\bsdsrho}{{\bs\to D_s^-\rho^+}}
\newcommand{\bsdsstrho}{{\bs\to D_s^{\ast-}\rho^+}}
\newcommand{\bsdsSTrho}{{\bs\to D_s^{(\ast)-}\rho^+}}
\newcommand{\bsjpsieta}{{\bs\to\jpsi\,\eta}}
\newcommand{\bsjpsietaP}{{\bs\to\jpsi\,\eta^{(')}}}
\newcommand{\bsjpsietap}{{\bs\to\jpsi\,\eta^{'}}}

\newcommand{\bsjpsifz}{{\bs\to\jpsi\,\fz}}
\newcommand{\bspipi}{{\bs\to\pi^+\pi^-}}
\newcommand{\bskpi}{{\bs\to K^-\pi^+}}
\newcommand{\bskk}{{\bs\to K^+K^-}}
\newcommand{\bskzkz}{{\bs\to K^0\bar K^0}}

\newcommand{\fss}{{\left(90.1^{+3.8}_{-4.0}\pm0.2\right)\%}}
\newcommand{\fs}{{\left(7.3^{+3.3}_{-3.0}\pm0.1\right)\%}}
\newcommand{\ff}{{\left(2.6^{+2.6}_{-2.5}\right)\%}}

\newcommand{\mbs}{{\left(5364.4\pm1.3\pm0.7\right)\mmev}}
\newcommand{\mbsst}{{\left(5416.4\pm0.4\pm0.5\right)\mmev}}

\newcommand{\fl}{{1.05^{+0.08}_{-0.10}{}^{+0.03}_{-0.04}}}

\newcommand{\bfbstojpsieta}{{(3.32\pm0.87({\rm stat.}){}^{+0.32}_{-0.28}({\rm syst.})\pm0.42(f_s))\times10^{-4}}}
\newcommand{\bfbstojpsietaprime}{{(3.1\pm1.2({\rm stat.})^{+0.5}_{-0.6}({\rm syst.})\pm0.4(f_s))\times 10^{-4}}}

\newcommand{\bfbstokk}{{(3.8{}^{+1.0}_{-0.9}({\rm stat.})\pm0.5({\rm syst.})\pm0.5(f_s))\times 10^{-5}}}
\newcommand{\bfbstokpi}{{2.6\times10^{-5}}} 
\newcommand{\bfbstopipi}{{1.2\times10^{-5}}} 
\newcommand{\bfbstokzkz}{{6.6\times 10^{-5}}} 

\newcommand{\bfbstodsstdsst}{{(3.1^{+1.2}_{-1.0}({\rm stat.})\pm0.8({\rm syst.}))\%}}
\newcommand{\bfbstodsSTdsST}{{(6.9^{+1.5}_{-1.3}({\rm stat.})\pm1.9({\rm syst.}))\%}}
\newcommand{\bfbstodsstds}{{(2.8^{+0.8}_{-0.7}({\rm stat.})\pm0.7({\rm syst.}))\%}}
\newcommand{\bfbstodsds}{{(1.0^{+0.4}_{-0.3}({\rm stat.})^{+0.3}_{-0.2}({\rm syst.}))\%}}
\newcommand{\dGoG}{{(14.7^{+3.6}_{-3.0}({\rm stat.}){}^{+4.4}_{-4.2}({\rm syst.}))\times10^{-2}}}

\newcommand{\bfbstojpsifz}{{1.63\times10^{-4}\textrm{~(at 90\% C.L.)}}} 

\title{$\pmb{\bs}$ Decays at Belle }

\ShortTitle{$\bs$ Decays at Belle}

\author{\speaker{Remi Louvot}\\
  (On behalf of the Belle collaboration)\\
  Laboratoire de Physique des Hautes \'Energies,\\
  \'Ecole Polytechnique F\'ed\'erale de Lausanne~(EPFL), Lausanne, Switzerland\\
  E-mail: \email{remi.louvot@epfl.ch}}

\abstract{

  The large data sample recorded with the Belle detector at the
  $\FiveS$ energy provides a unique opportunity to study the poorly-known $\bs$ meson.
  Several analyses, made with a data sample representing an integrated luminosity of 23.6~$\invfb$, are presented.
  We report the study of the large-signal $\bs\to D_s^{(\ast)-}h^+$ ($h^+=\pi^+,\rho^+$) decays
  including the first observations of  $\bsdsstpi$ and $\bsdsSTrho$.
  In addition, several results on $CP$-eigenstate $\bs$ decays are described.
  These include the study of the $\bsjpsietaP$ and $\bsjpsifz$ decays, the charmless $\bskk$, $\bspipi$ and $\bs\to\KS\KS$ decays
  and the simultaneous fit of the three $\bs\to\dsSTdsST$ modes from which $\dG/\G$ is extracted.
  The preliminary measurement of $\BR(\bsjpsifz)<\bfbstojpsifz$ is presented for the first time.

 ~

~

1 November 2010

LPHE Note 2010-06}

\FullConference{
  Flavor Physics and CP Violation - FPCP 2010\\
  May 25-29, 2010\\
  Turin, Italy}

\begin{document}

\section*{Introduction}

The Belle experiment \cite{NIMA_479_117}, located at the interaction point of
the KEKB asymmetric-energy $e^+e^-$ collider \cite{NIMA_499_1},
was designed for the study of $B$ mesons\footnote{The notation ``$B$'' refers either to a $B^0$ or a $B^+$.
  Moreover, charge-conjugated states are implied everywhere.}
produced in $e^+e^-$ annihilation at a center-of-mass (CM) energy corresponding to the mass of
the $\FourS$ resonance ($\sqrt s\approx10.58~\gev$).
After having recorded an unprecedented sample of $\sim700$ millions of $B\bar B$ pairs,
the Belle collaboration started to record collisions at higher energies,
opening the possibility to study other particles, like the $\bs$ meson.
Up to now, a data sample of integrated luminosity of $\lint=(23.6\pm0.3)\invfb$ (out of a total of $120~\invfb$) has been analyzed at the energy of the $\FiveS$ resonance
($\sqrt s\approx10.87~\gev$).

Since the $\FiveS$ resonance is just above the $\bs\bar\bs$ threshold, it was naturally
expected that the $\bs$ meson could be studied with $\FiveS$ data as well as
the $B$ mesons are with $\FourS$ data.
The large potential of such $\FiveS$ data was quickly confirmed
\cite{PRL_98_052001,PRD_76_012002} with the 2005 engineering run
representing 1.86~$\invfb$.
The main advantage with respect to the hadronic colliders is
the possibility of measurements of absolute branching fractions.
However, the abundance of $\bs$ mesons in $\FiveS$ hadronic events has to be
precisely determined.
Above the $e^+e^-\to u\bar u, d\bar d, s\bar s, c\bar c$ continuum events,
the $e^+e^-\to b\bar b$ process can produce different kinds of final states involving
a pair of non-strange $B$ mesons \cite{PRD_81_112003} ($B^{\ast}\bar B^{\ast}$,
$B^{\ast}\bar B$, $B\bar B$, $B^{\ast}\bar B^{\ast}\pi$, $B^{\ast}\bar B\pi$,
$B\bar B\pi$, $B\bar B\pi\pi$ and $B\bar B\gamma$),
a pair of $\bs$ mesons ($\bsst\barbsst$, $\bsst\bar\bs$ and $\bs\bar\bs$),
or final states involving a lighter bottomonium resonance below the open-beauty
threshold \cite{PRL_100_112001}.
The $B^{\ast}$ and $\bsst$ mesons always decay by emission of a photon.
The total $e^+e^-\to b\bar b$ cross section at the $\FiveS$ energy was measured
to be $\sigma_{b\bar b}=(302\pm14)$~pb
\cite{PRL_98_052001,PRD_75_012002} and the fraction of $\bs$ events to be\footnote{The branching-fraction values for $\bsdspi$ and those in Sections \ref{sec:jpsi} and \ref{sec:bshh} 
  are calculated with $f_s=(19.5^{+3.0}_{-2.3})\%$, also provided in Ref.~\cite{PLB_667_1}.}
$f_s=\sigma(e^+e^-\to\bsSTbsST)/\sigma_{b\bar b}=(19.3\pm2.9)$~\% \cite{PLB_667_1}.
The dominant $\bs$ production mode, $b\bar b\to\bsst\barbsst$,
represents $f_{\bsst\barbsst}=\fss$ of the $b\bar b\to\bsSTbsST$ events, as measured with $\bsdspi$ events (next Section).

For all the exclusive modes presented here, the $\bs$ candidates are fully reconstructed
from the final-state particles.
From the reconstructed four-momentum in the CM, $(E_{\bs}^{\ast},\pmb{p}_{\bs}^{\ast})$,
two variables are formed:
the energy difference $\deltae=E_{\bs}^{\ast}-\sqrt s/2$ and the
beam-constrained mass $\mbc=\sqrt{s/4-\pmb{p}_{\bs}^{\ast2}}$.
The signal coming from the dominant $e^+e^-\to\bsst\barbsst$ production mode
is extracted from a two-dimensional fit performed on the distribution of these
two variables.
The corresponding branching fraction is then extracted using the total efficiency (including sub-decay branching fractions)
determined with Monte-Carlo (MC) simulations, $\sum\varepsilon\BR$,
and the number of $\bs$ mesons produced via the $e^+e^-\to\bsst\barbsst$ process, $N_{\bs}=2\times\lint\times\sigma_{b\bar b}\times f_s\times f_{\bsst\bar\bsst}=(2.5\pm0.4)\times10^6$.

\section{Dominant CKM-favored $\pmb{\bs}$ Decays}
      
We report the measurement of exclusive $\bs\to D_s^{(\ast)-}h^+$ ($h^+=\pi^+$ or $\rho^+$) decays \cite{PRL_102_021801,PRL_104_231801} 
which is an important milestone in the study
of the poorly-known decay processes of the $\bs$ meson.
These modes are expected to produce an abundant signal because of their relatively large predicted branching fractions \cite{PLB_318_549,PRD_78_014018}
and their clean signatures: four charged tracks and up to two photons.
The leading amplitude for the four $\bsdsSTpi$ and $\bsdsSTrho$ modes
is a $b\to c$ tree diagram of order $\lambda^2$
(in the Wolfenstein parametrization~\cite{PRL_51_1945}
of the CKM quark-mixing matrix~\cite{PRL_10_531,PTP_49_652}) with a spectator $s$ quark.
Besides being interesting in their own right, such measurements, if precise enough, 
can be of high importance for the current and forthcoming hadron collider experiments. 
It was for example recently pointed out~\cite{hepex_0912_4179} that the search 
for the very rare decay $\bs\to\mu^+\mu^-$, which 
has a branching fraction very sensitive to New Physics contributions, will be 
systematically limited at LHCb by the poor knowledge of $\bs$ production, in case 
New Physics will enhance the decay probability by no more than a factor 3 
above the Standard Model expectation.

\begin{figure}[!b]
  \centering
\begin{minipage}{0.8\linewidth}
\centering
  \includegraphics[width=\linewidth]{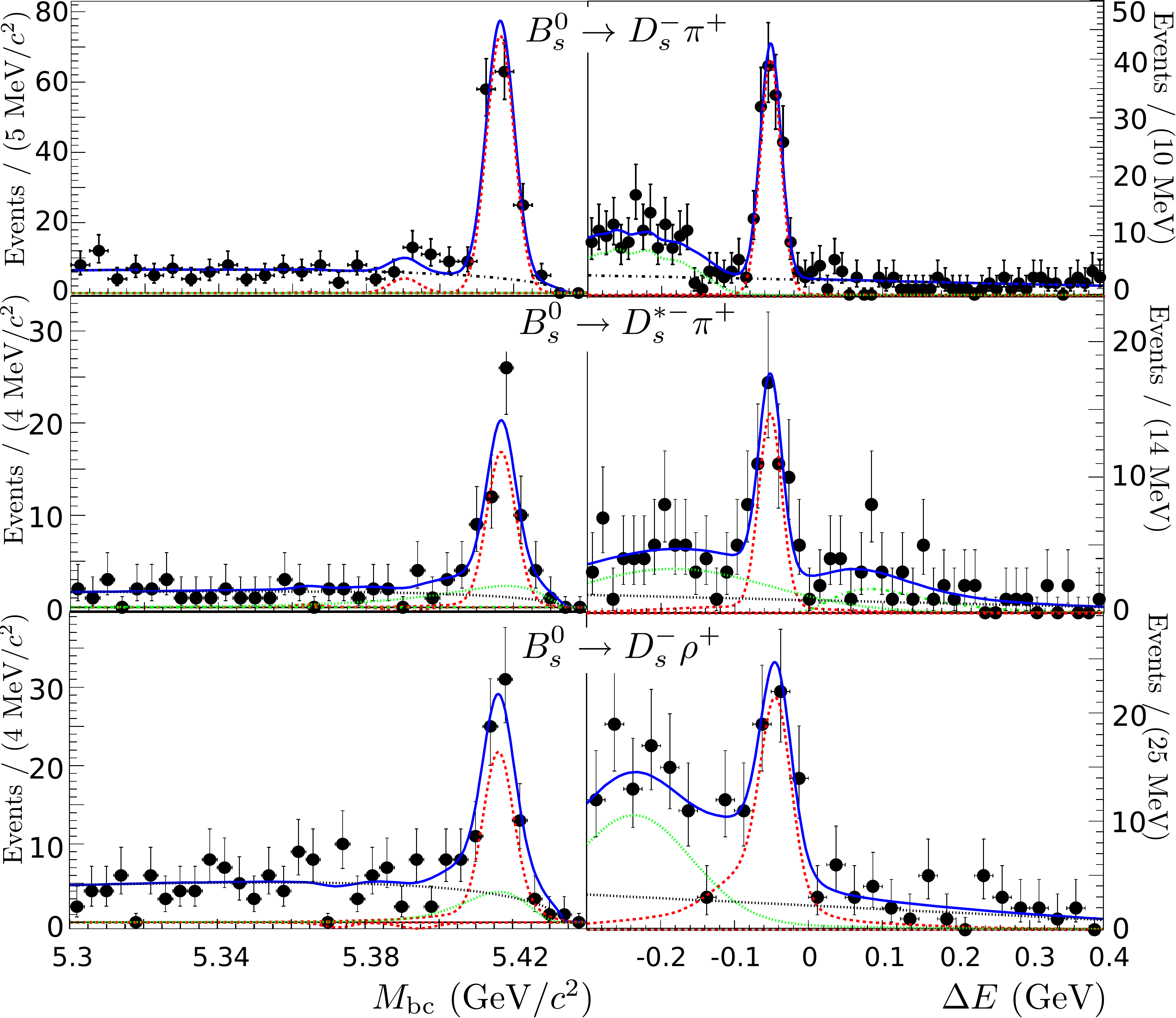}
\end{minipage}

  \caption{\label{fig2} Left: $\mbc$ distributions for the $\bsdspi$ (top) $\bsdsstpi$ (middle) and $\bsdsrho$ (bottom) candidates
    with $\deltae$ restricted to the $\bsst\barbsst$ signal region.
    Right: $\deltae$ distributions with $\mbc$ restricted to the $\bsst\barbsst$ signal region.
    The black- (green-) dotted line represents the continuum (peaking) background,
    while the red-dashed curves are the signal shapes.
    The larger one is the signal in the $\bsst\barbsst$ kinematic region and the two others,
    which are very close to 0, are the signals in the two other $\bs$ production modes ($\bsst\bar\bs$ and $\bs\bar\bs$).  }
\end{figure}

In addition, polarization measurements of $B$ decays have become of high interest since the observation
of a surprisingly large transverse polarization in $B\to\phi K^{\ast}$ decays by Belle and BaBar~\cite{PRL_91_171802,PRL_91_201801}.
The relative strengths of the longitudinal and transverse states can be
measured with an angular analysis of the decay products.
In the helicity basis, the expected $\bsdsstrho$ differential decay width is proportional to 
$$\frac{{\rm d}^2\Gamma(\bsdsstrho)}{{\rm d}\cdsst {\rm d}\crho}\propto 4f_L\ssqdsst\csqrho+(1-f_L)(1+\csqdsst)\ssqrho\,,$$
where $f_L=|H_0|^2/\sum_{\lambda}|H_{\lambda}|^2$ is the longitudinal polarization fraction, 
$H_{\lambda}$ ($\lambda=\pm1,0$) are the helicity amplitudes, and 
$\tdsst$ ($\trho$)  is the helicity angle of the $D_s^{\ast-}$ ($\rho^+$)
defined as the supplement of the angle between the $\bs$ and the $\ds$ ($\pi^+$) momenta
in the $\dsst$ ($\rho^+$) frame.

\begin{figure}[!t]
  \centering
  \includegraphics[width=0.8\linewidth]{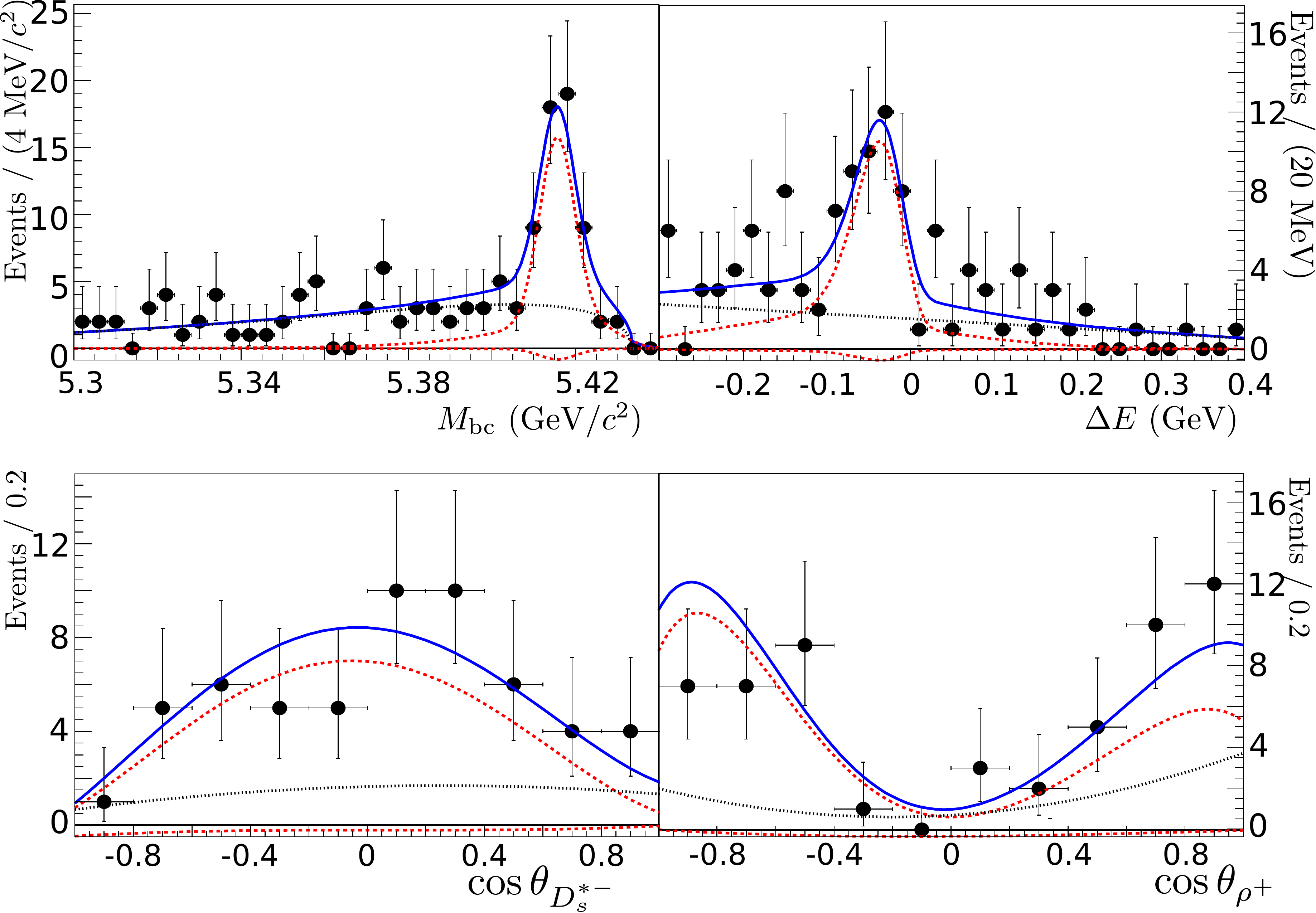}
  \caption{\label{fig3}     
    Fit of the $\bsdsstrho$ candidates.
    Top: $\mbc$ and $\deltae$ distributions, similarly to Fig.~1. 
    Bottom: helicity distributions of the $D_s^{\ast -}$ (left) and $\rho^+$ (right) with $\mbc$
    and $\deltae$ restricted to the $\bsst\barbsst$ kinematic region.
    The black-dotted line represents the background, while the two red-dashed curves are the signal.
    The large (small) signal shape corresponds to the longitudinal (transverse) component.
  } 
\end{figure}

The $\ds$ mesons are reconstructed via three modes :
$\ds\to\phi(\to K^+K^-)\pi^-$, $\ds\to K^{\ast0}(\to K^+\pi^-)K^-$ and
$\ds\to\KS(\to\pi^+\pi^-)K^-$.
Based on the ratio of the second and the zeroth Fox-Wolfram moments \cite{PRL_41_1581}, $R_2$,
the continuum events are efficiently
rejected by taking advantage of the difference between their event geometry (jet like, high $R_2$) and
the signal event shape (spherical, low $R_2$).
The $\bsdspi$ and $\bsdsstpi$ ($\bsdsrho$ and $\bsdsstrho$) candidates with $R_2$ smaller than 0.5 (0.35) are kept for further analysis.
A best candidate selection, based on the intermediate-particle reconstructed masses, is then implemented in order to keep only one $\bs$ candidate per event.
The $\mbc$ and $\deltae$ distributions of the selected $\bs$ candidates for the three $D_s^-$
modes are shown in Figs.~\ref{fig2} and \ref{fig3}, where the various components of the probability density function (PDF) used for the fit are described.
The $\bsdsstrho$ candidates are observed with two additional variables, $\cdsst$ and $\crho$,
which are the cosines of the helicity angles defined above.
They are needed for the measurement of the longitudinal polarization fraction, $f_L$.

Table~\ref{summary} presents a summary of the numerical results obtained for the $\bsdsSTpi$ and $\bs\to$ $D_s^{(\ast)-}\rho^+$ modes.
The different sources of systematic uncertainties affecting the measurements are identified and quoted as a second error. Our results on the $\bs$ decays are consistent with theoretical predictions \cite{PLB_318_549,PRD_78_014018} and with existing measurements (Table~\ref{summary}).

\begin{table}[!ht]
  \centering
    \centering    
    \renewcommand{\arraystretch}{1.3}
    
    \begin{tabular}{lc@{\hspace{0.4cm}}c@{\hspace{0.4cm}}c@{\hspace{0.4cm}}c|@{\hspace{0.4cm}}c}
      Mode         & $N_{\bsst\barbsst}$    & $S$          & $\varepsilon$ ($10^{-3}$)      & $\BR$ ($10^{-3}$)                & $\BR$ World average ($10^{-3}$)\\
      \hline
      $\bsdspi$    & $145^{+14}_{-13}$      & $21\sigma$   & $15.8$                         & $3.7^{+0.4}_{-0.3}\pm0.4\pm0.5$  & $3.2\pm0.9$~\cite{PLB_667_1}\\
      $\bsdsstpi$  & $53.4^{+10.3}_{-9.4}$  & $7.1\sigma$  & $9.13$                         & $2.4^{+0.5}_{-0.4}\pm0.3\pm0.4$  & First measurement\\
      $\bsdsrho$   & $92.2^{+14.2}_{-13.2}$ & $8.2\sigma$  & $4.40$                         & $8.5^{+1.3}_{-1.2}\pm1.1\pm1.3$  & First measurement\\
      $\bsdsstrho$ & $77.8^{+14.5}_{-13.4}$ & $7.4\sigma$  & $2.67$                         & $11.9^{+2.2}_{-2.0}\pm1.7\pm1.8$ & First measurement\\
    \end{tabular}


   \begin{tabular}{cc|c}
      Observable          &This work & World average\\
      \hline
      $m(\bs)$            &$\mbs$    & $(5366.4\pm1.1)\mmev$~\cite{PDG07}\\
      $m(\bsst)$          & $\mbsst$ & $(5411.7\pm1.7)\mmev$~\cite{PRL_96_152001}\\
      $f_{\bsst\barbsst}$ & $\fss$   & $(93^{+7}_{-9})\%$~\cite{PRD_76_012002}\\
      $f_{\bsst\bar\bs}$  & $\fs$    & First measurement\\
      $f_{\bs\bar\bs}$    & $\ff$    & First measurement\\
      $f_L(\bsdsstrho)$   & $\fl$    & First measurement
    \end{tabular}

   \caption{\label{summary}
     Summary of the results for the four $\bsdsSTpi$ and $\bsdsSTrho$ modes \cite{PRL_102_021801,PRL_104_231801}.
     Top: signal yields in the $\bsst\barbsst$ production mode, $N_{\bsst\barbsst}$, 
     significances, $S$, including systematics,
     total signal efficiencies, $\varepsilon$ (including all sub-decay branching fractions), and
     branching fractions, $\BR$, where the uncertainty due to $f_s$ (third error) is separated from the others systematics (second error).
     The first error represents the statistical uncertainties.
     Bottom: other measurements obtained with the $\bsdspi$ analysis and $\bsdsstrho$ longitudinal polarization fraction.
     The world averages (made without the measurements presented here) are shown for comparison in the last column of the tables.
   }  
\end{table}

\section{Study of $\pmb{\bs\to J\!/}\!\mathbf{\psi\,\eta^{(')}}$ and Search for $\pmb{\bs\to J\!/}\!\mathbf{\psi}\,\pmb{f_0}$($\mathbf{980}$)}\label{sec:jpsi}
\begin{figure}[!ht]
  \centering
  \begin{minipage}{0.4\linewidth}
    \centering
    \includegraphics[width=\linewidth,height=4cm]{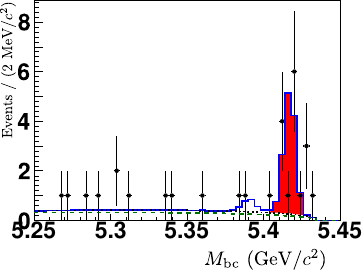}
  \end{minipage}~~~
  \begin{minipage}{0.4\linewidth}
    \centering
    \includegraphics[width=\linewidth,height=4cm]{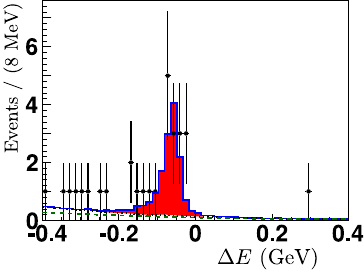}
  \end{minipage}
  \caption{\label{fig:jpsieta}$\mbc$ (left) and $\deltae$ (right) distributions, similarly to Fig.~1, 
    of the $\bsjpsieta$ candidates
    (points with error bars) and the fitted PDF (solid line).
    The sub-modes $\eta\to\gamma\gamma$ and $\eta\to\pi^+\pi^-\pi^0$, which are fitted separately, are summed in these plots.
    The green-dotted line (red region) represents the continuum (signal) component of the PDF.
    The small peak in the $\mbc$ plot is the $\bsst\bar\bs$ contribution, as the $\bsst\barbsst$ signal range in $\deltae$ overlaps with that of the $\bsst\bar\bs$ signal.}
\end{figure}

\begin{figure}[!ht]
  \centering
  \includegraphics[width=0.8\linewidth,height=7.5cm]{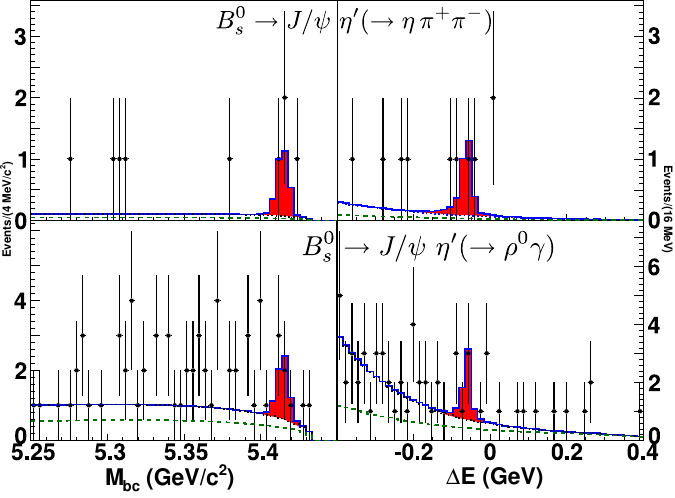}
  \caption{\label{fig:jpsietap}$\mbc$ (left) and $\deltae$ (right) distributions, similarly to Fig.~1, 
    of the $\bsjpsietap$ candidates
    (points with error bars) and the fitted PDF (solid line).
    The green-dotted line represents the continuum component of the PDF.
    The red region represents the signal component of the PDF.
  }
\end{figure}

$\bs$ decays to $CP$ eigenstates are important for $CP$-violation parameter measurements \cite{PRD_63_114015}.
Results about the first observation of
$\bsjpsieta$ and the first evidence for $\bsjpsietap$  are reported~\cite{hepex_0912_1434}.
The $\jpsi$ candidates are formed with oppositely-charged electron or muon pairs,
while $\eta$ candidates are reconstructed via the $\eta\to\gamma\gamma$ and $\eta\to\pi^+\pi^-\pi^0$ modes.
A mass (mass and vertex) constrained fit is then applied to the $\eta$ ($\jpsi$) candidates.
The $\eta^{'}$ candidates are reconstructed via the $\eta^{'}\to\eta\pi^+\pi^-$ and $\eta^{'}\to\rho^0\gamma$ modes, while the $\rho^0$ candidates are selected from $\pi^+\pi^-$ pairs.
If more than one candidate per event satisfies all the selection criteria, the one with the smallest fit residual is selected.
The main background is the continuum, which is reduced by requiring $R_2<0.4$. 
The combined $\mbc$ and $\deltae$ distributions are presented in Figs.~\ref{fig:jpsieta} ($\bsjpsieta$) and \ref{fig:jpsietap} ($\bsjpsietap$).
We obtain $\BR(\bsjpsieta)=\bfbstojpsieta$ and $\BR(\bsjpsietap)=\bfbstojpsietaprime$.
This is, respectively, the first observation ($7.3\sigma$) and the first evidence (3.8$\sigma$) for these modes.

The $\bsjpsifz$ mode is especially interesting for the hadron-collider experiments because it has only four charged tracks in its final state.
Recent calculations predict the ratio $$\rfphi=\frac{\BR(\bsjpsifz)\times\BR(\fz\to\pi^+\pi^-)}{\BR(\bs\to\jpsi\,\phi)\times\BR(\phi\to K^+K^-)}$$
to be $\approx0.2$ \cite{PRD_79_074024}.
From the CLEO analysis of $D_s^+\to\fz e^+\nu_e$, $\rfphi$ is estimated to be $0.42\pm0.11$ \cite{PRD_80_052009}.
From QCD estimates \cite{PRD_81_074001} and BES result of $\BR(\fz\to\pi^+\pi^-)$, $\rfphi\approx0.24$. 
Other predictions from generalized QCD factorization \cite{PRD_82_076006} are compatible with these estimates.

\begin{figure}[!t]
  \centering
  \includegraphics[width=0.4\linewidth,height=4cm]{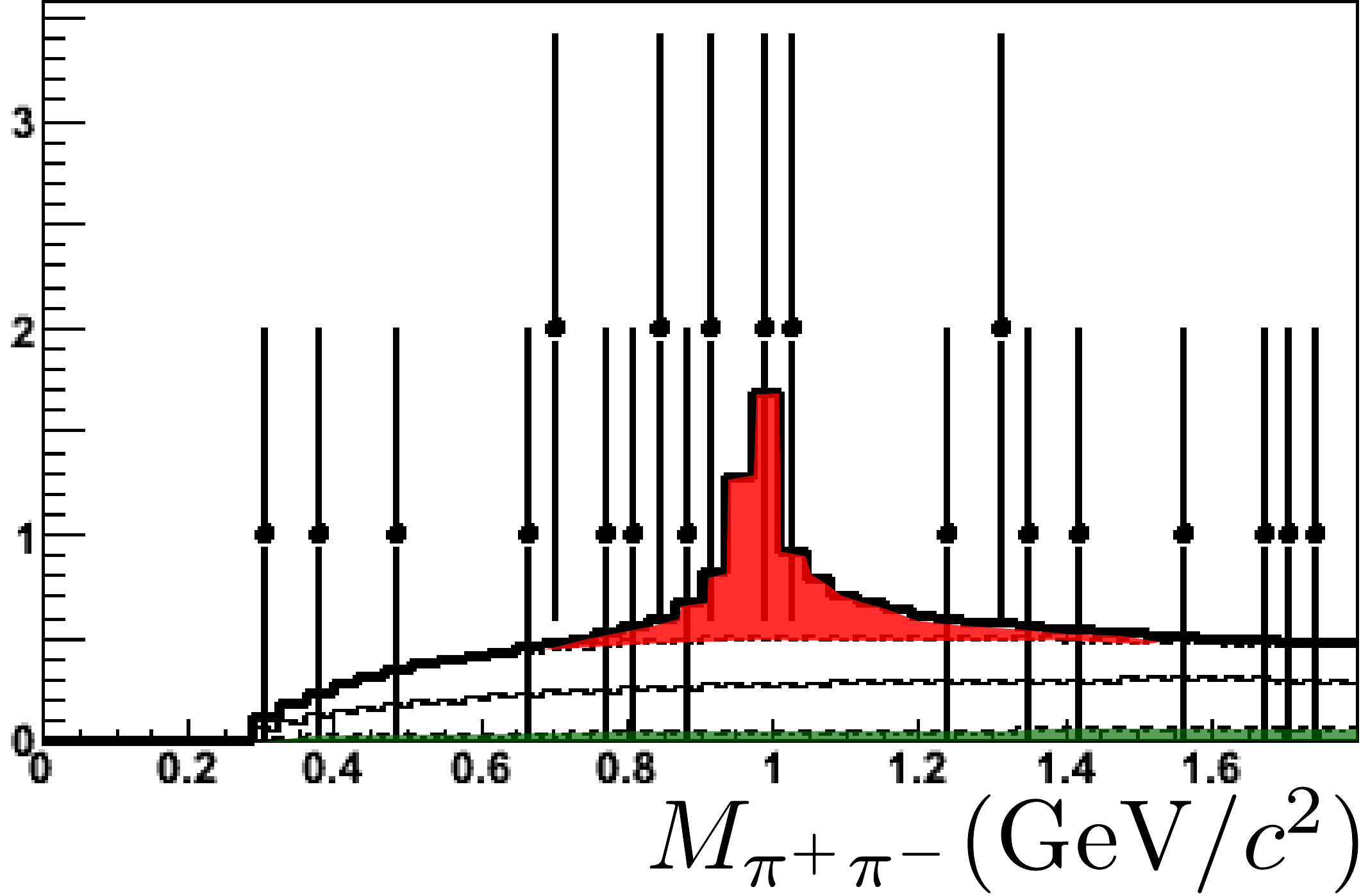}
  ~~
  \includegraphics[width=0.4\linewidth,height=4cm]{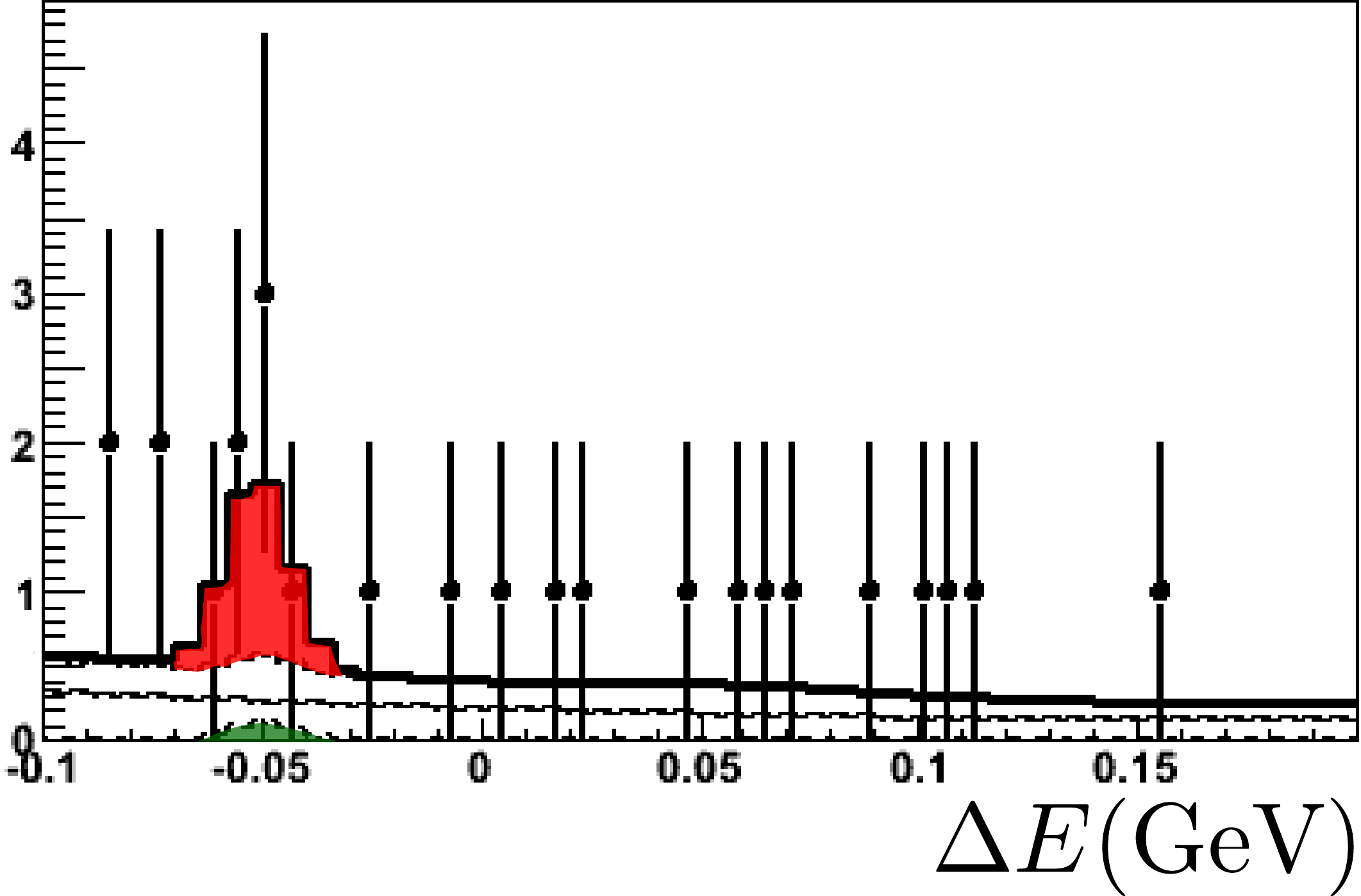}
  \caption{\label{fig:fz} $\fz$ mass (left) and $\deltae$ (right) distributions of the $\bsjpsifz$ candidates.
    The solid-black line is the total fitted PDF.
    The green region represents the contribution of the non-resonant $\bs\to\jpsi\,\pi^+\pi^-$, while the red region is the signal.
    The dotted-black curve is the contribution of the other $\bs\to\jpsi\,X$ modes.}
\end{figure}

With the same selection for the $\jpsi$ as described above, and the reconstruction of $\fz\to\pi^+\pi^-$ candidates,
the $\bsjpsifz$ signal is fitted using the energy difference, $\deltae$, and the $\fz$ mass, $M_{\pi^+\pi^-}$, distributions (Fig.~\ref{fig:fz}).
No significant signal ($6.0\pm4.4$ events, 1.7$\sigma$) is seen and we set the upper limit $$\BR(\bsjpsifz)\times\BR(\fz\to\pi^+\pi^-)<\bfbstojpsifz\,,$$
or, similarly, $$\rfphi<0.275\textrm{~~~(at 90\% C.L.)}$$ using our preliminary result of $\BR(\bs\to\jpsi\,\phi)$ \cite{hepex_0905_4345}.
These limits are clearly in the region of interest and an update using our full data sample (120$\invfb$) is very important.

\section{Observation of $\pmb{\bskk}$ and Searches for
  $\pmb{\bspipi}$, $\pmb{\bskpi}$ and $\pmb{\bs\to\KS\KS}$}\label{sec:bshh}

We present our results for the $\bskk$, $\bskpi$, $\bspipi$ and $\bs\to\KS\KS$ charmless decays \cite{PRD_82_072007}.
The $\bskk$ mode is particularly interesting because it can be used for the determination of
the CKM angle $\gamma$ \cite{PLB_459_306} and may be sensitive to
New Physics \cite{PRD_70_031502}.
The charged pion and kaon candidates are selected using charged tracks and identified with energy deposition, momentum and time-of-flight measurements.
The $\KS$ candidates are reconstructed via the $\KS\to\pi^+\pi^-$ decay, by selecting two oppositely-charged tracks matching various geometrical requirements \cite{phd_ffang}.
A likelihood based on a Fisher discriminant using 16 modified Fox-Wolfram moments \cite{PRL_91_261801} is implemented to reduce the continuum, which is the main source of background.

\begin{figure}[!ht]
  \centering
  \begin{minipage}{0.3\linewidth}
    \centering
    \includegraphics[height=4cm,width=\linewidth]{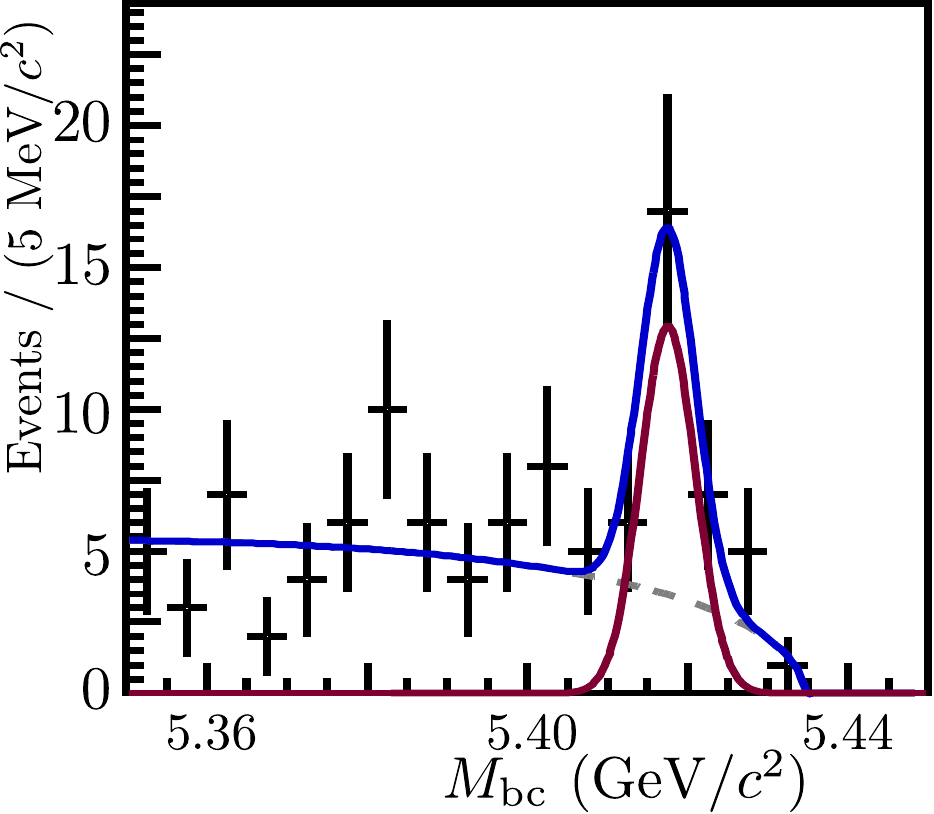}
  \end{minipage}~
  \begin{minipage}{0.3\linewidth}
    \centering
    \includegraphics[height=4cm,width=\linewidth]{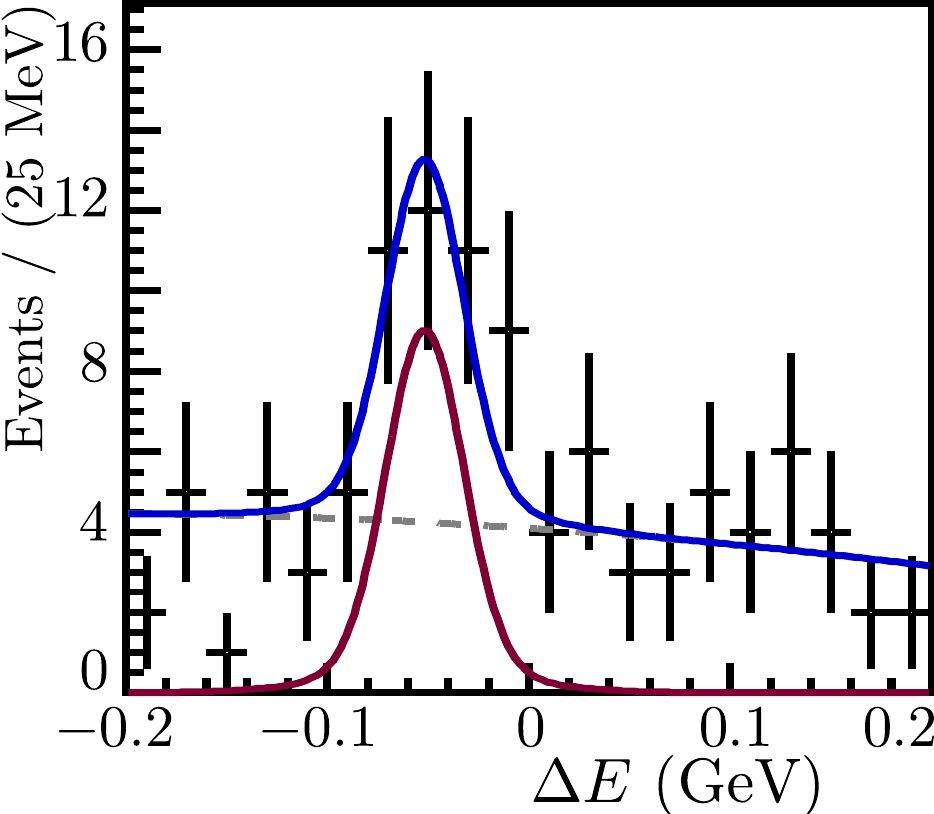}
  \end{minipage}~~~~
  \begin{minipage}{0.36\linewidth}
    \caption{\label{fig:kk}Distributions, similarly to Fig.~1, of the $\bs\to K^+K^-$ candidates
      and the fitted PDF (solid blue line).
      The solid-red and the dotted-grey curves represent the signal and the continuum
      component of the PDF, respectively.}
  \end{minipage}
\end{figure}

We do observe a 5.8$\sigma$ excess of $24\pm6$ events in the $\bsst\barbsst$ region for the $\bskk$ mode (Fig.~\ref{fig:kk}).
The branching fraction $\BR(\bskk)=\bfbstokk$ is derived.
However, no significant signal is seen for the other modes.
Including the systematics uncertainties, we set the following upper limits at 90\% confidence level:
$\BR(\bspipi)$ $<\bfbstopipi$, $\BR(\bskpi)<\bfbstokpi$ and, assuming $\BR(\bs\to K^0\bar K^0)=2\times\BR(\bs\to\KS\KS)$, $\BR(\bskzkz)<\bfbstokzkz$.
The later is the first limit set for the $\bskzkz$ mode.
All the other values are compatible with the CDF results \cite{PRL_97_211802,PRL_103_031801}.

\section{Study of $\pmb{\bs\to\dsSTdsST}$ and Measurement of $\pmb{\dG/\G}$}

\begin{figure}[!ht]
  \centering
  \begin{minipage}{0.72\linewidth}
    \centering
    \includegraphics[width=\linewidth]{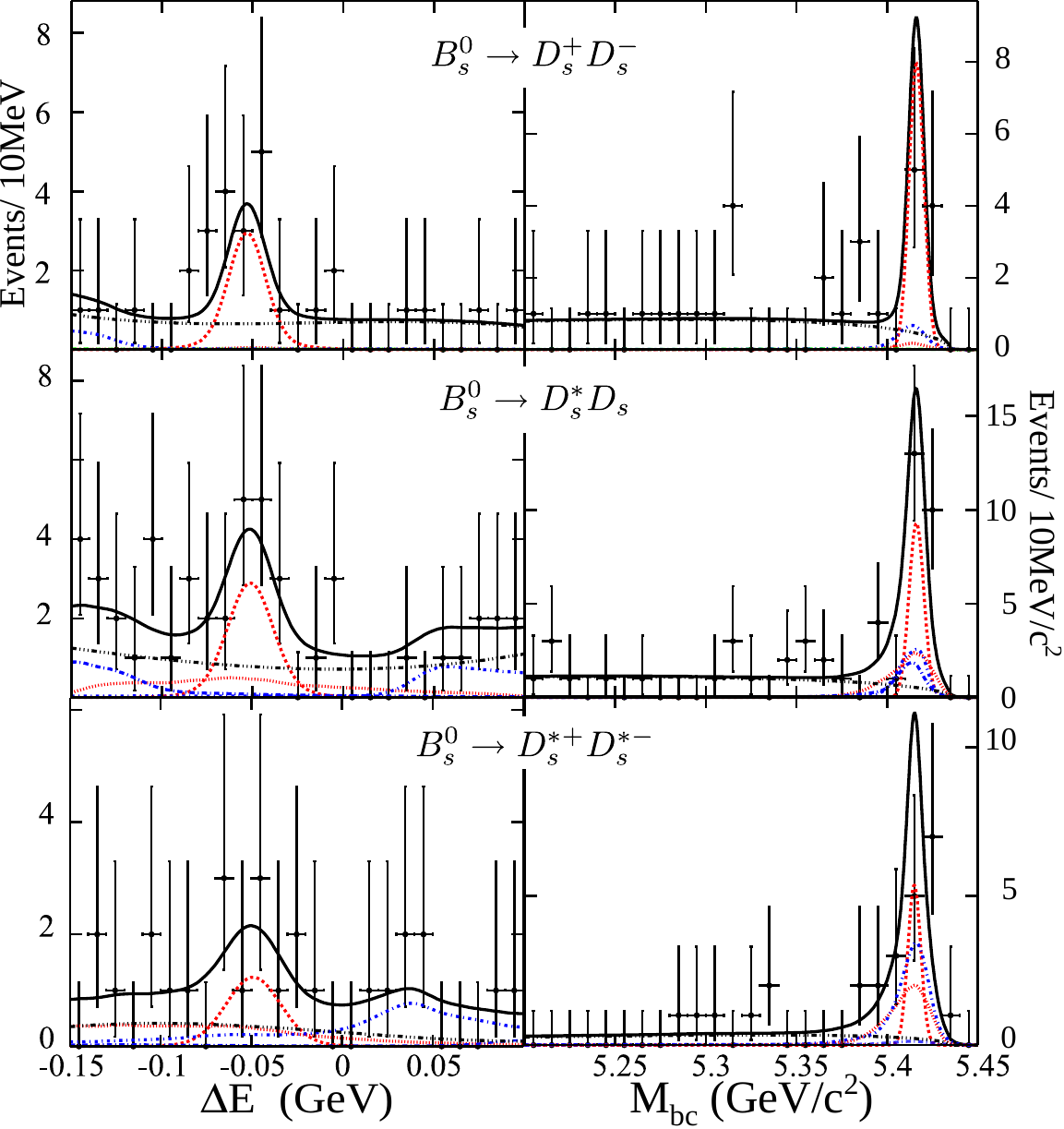}
  \end{minipage}
  ~~
  \begin{minipage}{0.25\linewidth}
    \caption{\label{fig:dsstdsst}$\deltae$ (left) and $\mbc$ (right) distributions, similarly to Fig.~1,
      of the $\bs\to D_s^+D_s^-$ (top), $\bs\to D_s^{\ast\pm}D_s^{\mp}$ (middle) and $\bs\to D_s^{\ast+}D_s^{\ast-}$ (bottom) candidates,
      together with the fitted PDF.
      Except the continuum background component, which is shown by the black dashed-dotted curve, all the other contributions are peaking in $\mbc$.
      The correct (wrong) combination signal, shown by the peaking (smooth) red dashed curve and the cross-feed components, shown by the blue dashed-dotted curve are well separated in $\deltae$.
    }
  \end{minipage}
\end{figure}

We finally report the results from our analysis of the $\bs\to\dsSTdsST$ decays \cite{hepex_1005_5177}.
These modes are $CP$ eigenstates \emph{and} CKM favored  ($b\to c\bar cs$ transition of order $\lambda^2$).
In the heavy-quark limit, they are $CP$ even and dominate $\Delta\Gamma$ \cite{PLB_316_567}.
The relative width difference of the $\bs-\bar\bs$ system can be obtained from the relation
\begin{equation}\label{eq:dg}\frac{\dG}{\G}=\frac{2\times\BR(\bs\to\dsSTdsST)}{1-\BR(\bs\to\dsSTdsST)}\,.\end{equation}
In order to reconstruct the $\bs\to\dsSTdsST$ candidates, we form $\ds$ candidates from 6 modes: $\ds\to\phi\pi^-$, $\ds\to K^{\ast0}K^-$, $\ds\to\KS K^-$, $\ds\to\phi\rho^-$, $\ds\to K^{\ast0}K^{\ast-}$ and $\ds\to\KS K^{\ast-}$.
Only one candidate per event is selected using $M(\ds)$ and $M(\dsst)-M(\ds)$ informations.
The same likelihood as in the previous Section, based on modified Fox-Wolfram moments \cite{PRL_91_261801}, is used to reject 80\% of the continuum events, while 95\% of the signal is kept.
The $\deltae$ and $\mbc$ distributions for each of the three $\bs\to\dsSTdsST$ modes are fitted simultaneously.
The signal PDF is made of two components studied with signal MC simulations: 
the correctly reconstructed candidates and the wrong combinations in which a non-signal track (photon) is included in place of a true daughter track (photon).
In addition the so-called cross-feed contributions are included:
a $\dsstds$ ($\dsstdsst$) event can be selected as a $\dsds$ ($\dsstds$) candidate with a lower energy because one photon is missing; 
conversely, a $\dsds$ ($\dsstds$) candidate can be reconstructed as a $\dsstds$ ($\dsstdsst$) candidate with an additional photon, hence its energy larger than expected.

\begin{table}[!ht]
  \centering
  \begin{tabular}{c|cc|cc}
    Mode              &$N_{\rm sig.}$      &$S$        &$\BR$             &$\BR$ World Average\\  
    \hline
    $\bs\to\dsstdsst$ &$4.9^{+1.9}_{-1.7}$ &3.2$\sigma$&$\bfbstodsstdsst$ &First evidence\\
    $\bs\to\dsstds$   &$9.2^{+2.8}_{-2.4}$ &6.6$\sigma$&$\bfbstodsstds$   &First observation\\
    $\bs\to\dsds$     &$8.5^{+3.2}_{-2.6}$ &6.2$\sigma$&$\bfbstodsds$     &$(1.04^{+0.37}_{-0.34})\%$ \\
    \hline
    $\bs\to\dsSTdsST$ &$22.6^{+4.7}_{-3.9}$&           &$\bfbstodsSTdsST$ &$(4.0\pm1.5)\%$\\
  \end{tabular}
  \caption{\label{tab:dsstdsst} Signal event yields, $N_{\rm sig.}$, significances, $S$, including systematics and branching fractions, $\BR$, for the three $\bs\to\dsSTdsST$ modes and their sum.
    The world averages, performed from other existing measurements \cite{PLB_486_286,PRL_100_021803,PRL_102_091801}, are those reported in Ref.~\cite{PDG10}.}
\end{table}

The fit results can be seen in Fig.~\ref{fig:dsstdsst} while the numerical values are reported in Table~\ref{tab:dsstdsst}.
With Eq.~(\ref{eq:dg}), we extract $$\frac{\dG}{\G}=\dGoG\,.$$
This value is in agreement with the SM expectations \cite{JHEP_06_072} and with the results from ALEPH, $(25^{+21}_{-14})\%$ \cite{PLB_486_286}, 
D\O, $(7.2\pm3.0)\%$ \cite{PRL_102_091801}, and
CDF\footnote{This is a measurement of $\Delta\Gamma_s/\Gamma_s=(\dG\cos\phi)/\G$, $\phi$ being the $CP$-violating phase (assumed to be negligible).}, $(12^{+9}_{-10})\%$ \cite{PRL_100_121803}.
With only 23 fully-reconstructed signal events, our measurement is already competitive with the Tevatron values.
\section*{Conclusion}
We presented new results on $\bs$ decays obtained from 23.6 $\invfb$ of $\FiveS$ data recorded by the Belle detector.
While modes with large statistics can provide precise measurements of branching fractions and $\bsST$ properties, 
first observations of several $CP$-eigenstate $\bs$ decays are a confirmation
of the large potential of our 120$\invfb$ $e^+e^-\to\FiveS$ data sample and advocate an ambitious $\bs$ program at super-$B$ factories.

\bibliography{bib}{}
\bibliographystyle{JHEP}   
\end{document}